\documentclass[letterpaper]{article} 
\usepackage{aaai2026}  
\usepackage{tabularx}
\nocopyright

\usepackage{times}  
\usepackage{helvet}  
\usepackage{courier}  
\usepackage[hyphens]{url}  
\usepackage{graphicx} 

\usepackage{natbib}  
\usepackage{caption} 
\usepackage[most]{tcolorbox}
\usepackage{enumitem}
\usepackage{xcolor}
\usepackage{booktabs}
\usepackage{array}
\usepackage{tikz}
\usepackage{comment}
\usepackage{booktabs}
\usepackage{tabularx}
\usepackage{pgfplots}
\usepackage{subcaption}
\usepackage{tikz}
\usepackage[most]{tcolorbox}
\usepackage{adjustbox}
\usepackage{xcolor}
\usepackage{float} 

\usepackage{ragged2e}

\newcolumntype{Y}{>{\RaggedRight\arraybackslash}X}

\newcommand{\legacy}{\textsc{Steve}}
\newcommand{\newtool}{\textsc{People Tool}}

\newlist{surveyitems}{itemize}{1}
\setlist[surveyitems,1]{
    label=\textbullet,
    leftmargin=*,
    itemsep=1pt,
    topsep=1pt,
    parsep=0pt,
    partopsep=0pt
}

\usetikzlibrary{positioning, arrows.meta, shapes.multipart, fit}
\pgfplotsset{compat=1.18}
\usetikzlibrary{arrows.meta, positioning, shapes.geometric, fit, backgrounds, calc}

\newcolumntype{Y}{>{\raggedright\arraybackslash}X}

\newlist{guideitems}{itemize}{1}
\setlist[guideitems]{
    label=\textbullet,
    leftmargin=*,
    itemsep=1pt,
    topsep=1pt,
    parsep=0pt,
    partopsep=0pt
}
\title{AI Adoption Across a Multinational Workforce: Sociotechnical Conditions for GenAI Acceptance in Human Resources}
 
\author{
    Dalia Ali\textsuperscript{\rm 1}\thanks{Equal contribution.}\thanks{Corresponding author: dalia.ali@tum.de},
    Maria Jos\'{e} Rodr\'{i}guez Vel\'{a}zquez\textsuperscript{\rm 1,*},
    Manoel Horta Ribeiro\textsuperscript{\rm 3},\\
    Vera Liao\textsuperscript{\rm 2},
    Orestis Papakyriakopoulos\textsuperscript{\rm 1}
}
\affiliations{
    \textsuperscript{\rm 1}Technical University of Munich\\
    \textsuperscript{\rm 2}University of Michigan\\
    \textsuperscript{\rm 3}Princeton University\\
}

\begin{document}
\maketitle

\begin{abstract}

Generative AI (GenAI) deployment in the workplace is accelerating rapidly. Nevertheless, questions of \textit{who adopts, who benefits, and who is left behind and why} are still understudied. In this paper, we investigate these dynamics in the context of a multinational tech company transitioning from a legacy Human Resources (HR) search system to a GenAI-supported system, analyzing search log data, survey data (\textit{n=25}), and ten semi-structured interviews. Our findings show that adoption depended on the fit between the GenAI system's design assumptions and employees' work positionalities (role, spoken language, tenure). Further, we find that employees' trust in GenAI answers was built through source-checking, comparison among systems, and seeking input from colleagues or HR when in doubt. Our contribution is twofold. First, we provide empirical evidence of workplace GenAI adoption during a live organizational transition, showing that adoption is influenced by factors such as situational fit, search literacy, and trust calibration. It is also further shaped by knowledge conditions such as the system's content quality, employee training, and guidance. Second, we translate these findings into design considerations for inclusive deployment and adoption in high-stakes environments such as HR. We argue that organizations should design systems considering the role- and context-sensitive benefits they yield to different social groups. They also need to treat the organizational knowledge infrastructure as AI infrastructure to improve the accountability and usability of GenAI systems.

\end{abstract}

\section{Introduction}
Artificial Intelligence (AI) is becoming a general-purpose layer of organizational infrastructure, transforming processes and decision-making~\cite {Enholm2022}. Recent studies suggest that AI capabilities are improving rapidly, enabling systems to handle increasingly complex tasks with limited human intervention~\cite {Douglas2025,Kwa2025}. 
Improvements in AI capabilities can therefore affect organizations at multiple levels. At the task level, AI can increase productivity by automating routine cognitive work, supporting repetitive analytical tasks, and helping workers make faster or better-informed decisions \cite{brynjolfsson2024generative,lee2022ai,ali2025ai}. At the employee level, these systems may offer real-time assistance, feedback, and communication support, facilitating learning and increasing job satisfaction, especially for less-experienced employees~\cite{brynjolfsson2024generative,bhargava2020employees}. Last, at the organizational level, AI can make information more widely available across teams and support coordination between departments, reducing some of the silos and expertise imbalances that often limit organizational learning and collaboration \cite{dellacqua2025cybernetic}.

Yet, adoption lags behind technological progress. Recent industry reports show a gap between AI access, exploration, and exploitation. While workforce access to AI technologies is growing, few companies manage to move from the pilot to the production stage; 65\% of senior leaders struggle to tie productivity gains directly to AI adoption \cite{Deloitte2026,MITNANDA2025,EYPulse2025}. Previous work shows that adoption is shaped by different factors such as perceived usefulness and ease of use, social influence, data, technology infrastructure, organizational culture, ethics, and regulation \cite{Enholm2022,Venkatesh2022,davis1989usefulness}. However, these explanations do not fully capture the complex, interconnected, and dynamic environments in which technology is implemented; further, they offer limited insight into the complex situated user experiences and their fit with available technologies \cite{Lee2025,Shachak2019, Mogaji2024}.

To bridge this gap, we need to understand not only how employees use AI but also what drives different employees' decisions to adopt, reject, or selectively use AI during organizational transitions. Without such understanding, organizations risk failing to realize the value of their investments, missing the benefits of increased efficiency and performance, and reproducing uneven patterns of adoption among employees \cite{Enholm2022,Mlekus2020}. From the employee perspective, poor GenAI design may mean that a system does not support the task at hand, produces outputs that are difficult to evaluate or trust, or excludes some employees due to a lack of language fit, accessibility, and work-context needs \citep{gu2024dataassistants,tankelevitch2024metacognitive,liao2022match,weidinger2021ethical,budhwar2023hrmgenai, sarkar2022like}. Since barriers to GenAI adoption already exacerbate existing inequalities across workers, inclusive design is needed to ensure that efficiency benefits are not concentrated among some employees while others are left behind \cite{HumlumVestergaard2025}.

To examine AI adoption from a user-centered perspective and understand why some employees benefit from GenAI systems while others are left behind, we study a real organizational transition within a multinational tech company from an existing knowledge system for HR to one that incorporates GenAI capabilities into its retrieval process. The new GenAI-based search system relies on HR documents from the internal organizational repository to enable employees to access HR-related information faster and more accurately. Existing organizational leadership's willingness to implement an innovative AI solution and the availability of necessary technological and financial resources allow us to investigate actual sociotechnical factors that facilitate or impede adoption, rather than boundary conditions of technological availability. Specifically, we raise the following research question:

\begin{center}
\textbf{\textit{Why do users adopt or resist a GenAI-enhanced HR knowledge system during an organizational transition?}}
\end{center}

By combining one year of the employees' search log data from an existing knowledge search system, two weeks of data from the new GenAI-based search system, an exploratory survey, and ten semi-structured interviews with project team members and end-users, we show that AI adoption was not a simple migration from an old to a new system. Instead, it was characterized by selective use across parallel systems, situated fit across user roles, social contexts, and the technologies, user's existing GenAI search literacy, how users calibrated their trust in the new system, and organizational conditions such as content quality and user guidance. 

Our paper makes two main contributions: (1) we identify sociotechnical mechanisms through which workplace GenAI produces unequal benefits, showing that when systems are designed around assumptions of office-based access, digital confidence, organizational familiarity, and language fit, workers outside these assumptions are systematically less well served despite formal access; (2) we translate the findings into design considerations for inclusive and accountable workplace GenAI systems.

\section{Background}

\subsection{AI Adoption in Organizations}
As AI capabilities continue to advance rapidly, adoption remains uneven across both organizations and workers \citep{HumlumVestergaard2025, Deloitte2026, MITNANDA2025}. Explaining this variation requires examining adoption at multiple levels. At the organizational level, the Theory of Technology, Organization, and Environment (TOE) identifies technological, organizational, and environmental factors that shape whether organizations embrace new technologies \citep{tornatzky1990processes, baker2011technology}. However, TOE does not capture individual-level adoption decisions within organizations \citep{awa2017integrated, oliveira2011literature}. 

At the individual level, the Technology Acceptance Model (TAM) and the Unified Theory of Acceptance and Use of Technology (UTAUT) explain adoption through perceived usefulness, ease of use, performance expectancy, effort expectancy, social influence, and facilitating conditions \citep{davis1989usefulness, venkatesh2003unified}. These models have strong predictive power for traditional software, but were not designed for AI systems, which are probabilistic, opaque, and context-dependent in ways that fundamentally alter how users interact with technology \citep{Venkatesh2022}. Extensions of these models to AI domain have taken into account issues related to trust, transparency, and perceived risk, recognizing that users must cope with probabilistic outputs, limited explainability, and unpredictable behavior \citep{Venkatesh2022, Enholm2022, wolfe2025revisiting}. Nonetheless, the efficient use of GenAI technologies introduces additional interaction requirements that cannot be entirely accommodated by the previous trust models or adoption models \citep{tankelevitch2024metacognitive,zamfirescu2023johnny,sarkar2022like,weisz2023toward}.

Research on AI in the workplace consistently shows that these systems do not impact all workers equally. Several studies document how GenAI reshapes knowledge work \citep{Woodruff2024knowledge}, and how algorithmic systems create failure loops that systematically undermine certain types of work \citep{kawakami2024failure}. In HR research, most scrutiny has focused on algorithmic hiring \citep{raghavan2020mitigating, sanchezmonedero2020hiring}, and emotion AI in interviews has been shown to result in lower perceptions of justice among gender minorities \citep{ingber2025emotion}.

Despite evidence that AI affects workers unequally, existing adoption models still provide limited insight into how adoption unfolds through everyday interactions among users, technology, and organizational context \citep{Shachak2019, Mogaji2024}. They do not adequately explain how users learn, adapt, develop workarounds, or incorporate AI systems into practice \citep{Shachak2019};  nor how adoption is shaped by users' expertise, job role, task suitability, and organizational conditions \citep{Bankins2024}.

\subsection{User-centered considerations of AI systems} 

Addressing this gap requires moving beyond current adoption models and examining how users engage with AI systems in practice. User-centered AI adoption depends on more than access. How well employees benefit from the AI system depends on their ability to interpret its outputs, build trust, learn a new interaction logic, and rely on the organizational knowledge infrastructure underlying the system \citep{tankelevitch2024metacognitive, liao2022match, alavi2024genaiKM}. Although these factors are largely studied when people use systems, prior research does not take them into consideration when studying adoption dynamics. Rather, the focus lies more on defining isolated social and technical factors that shape adoption\citep{Venkatesh2022, wolfe2025revisiting}.

Research on Human-AI interactions shows that the level of trust in AI  systems is not a function of system accuracy; is also influenced by how users interact with the system \citep{liao2022match}. \citet{liao2022match} differentiate between trustworthy AI, which is related to system accuracy and fairness, and human trust, which requires perception and communication. They propose the MATCH model, in which trust arises from the interplay among model features, system affordances, trustworthiness cues, and users’ cognitive processing. \cite{bansal2021does} also show that explanations, such as supporting information features that promote trust in AI, do not automatically produce appropriate reliance; they can increase users' acceptance of AI recommendations even when those recommendations are incorrect. This means that reliance on the AI output needs to be studied empirically, rather than assumed based on the system's accuracy. 

A second factor is the interaction competencies GenAI demands of its users. Unlike traditional information retrieval systems, GenAI requires users to prompt, evaluate outputs, reformulate queries, and decide whether answers are reliable enough to act on \citep{tankelevitch2024metacognitive, zamfirescu2023johnny}. The development of such competence is not straightforward, as employees must develop proficiency in prompting, output evaluation, and knowing when GenAI is appropriate \citep{xia2026clever}. This is made more challenging because workers use GenAI within existing tasks rather than as a standalone activity \citep{xia2026clever}.

Another factor to consider is the organizational knowledge that GenAI systems draw from. Information systems are only as effective as the knowledge supporting them \citep{alavi2001knowledge}; and GenAI systems are no exception, since they retrieve and synthesize content from underlying content such as articles, metadata, and localized content instead of generating answers independently \citep{alavi2024genaiKM}. Recent work argues that GenAI may change how organizations manage knowledge by making retrieval and recombination more conversational and automated \citep{alavi2024genaiKM}.

Collectively, this literature identifies user-centered factors that shape how people use and evaluate GenAI systems. Given this, we investigate how the above and additional sociotechnical factors interact to shape adoption during a live HR organizational transition across a structurally diverse workforce, in which a GenAI system is introduced alongside a legacy system. 


\section{Methodology}
\begin{figure*}[htbp]
    \centering
    \includegraphics[width=\linewidth, height=6.3cm]{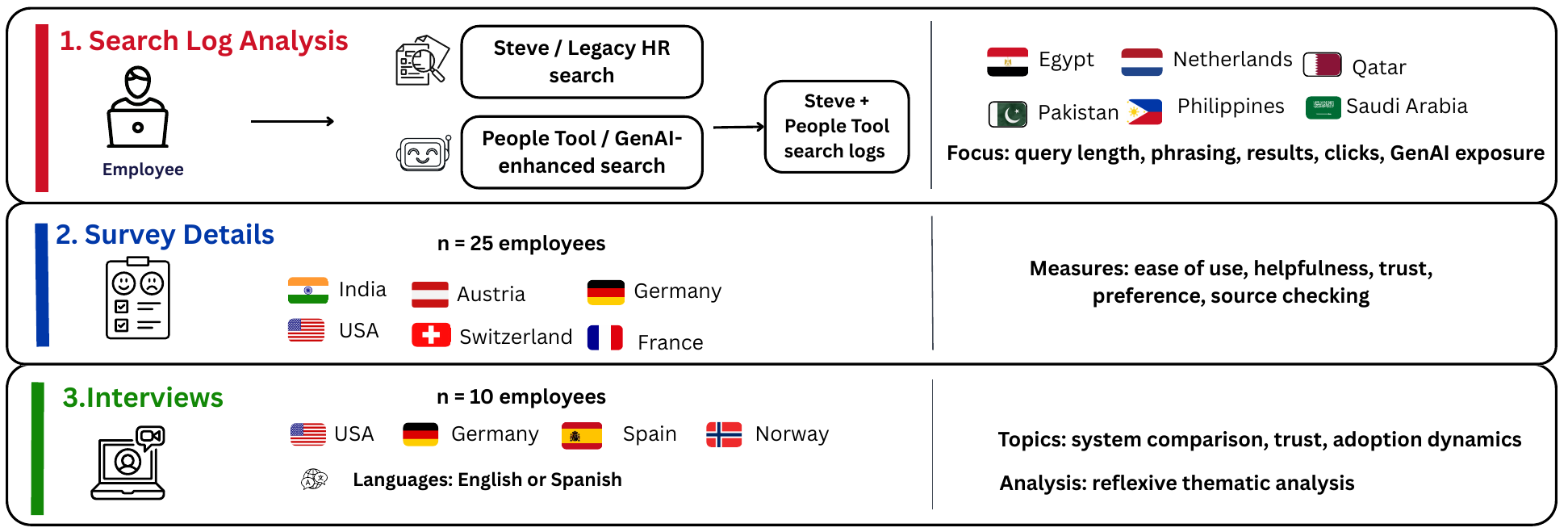}
    \caption{Sequential mixed-methods study design. Each phase informed the next, with search log patterns shaping the survey and survey responses shaping the interview protocol.}
    \label{fig:mixed-methods-design}
\end{figure*}

To understand how employees adopt or resist GenAI in organizational knowledge work, we conducted a mixed-method single-case study at a multinational technology organization with more than 300,000 employees worldwide. This case centered on the organization's introduction of \textsc{People Tool}\footnote{\textsc{People Tool} is a pseudonym for the newer GenAI-enhanced HR system.}, an HR platform with an embedded GenAI On Search feature, alongside the continued availability of its legacy HR search system, \textsc{Steve} \footnote{\textsc{Steve} is a pseudonym for the legacy HR knowledge system.}. \textsc{People Tool} operated as a retrieval-augmented system; employee queries returned either a GenAI-generated answer alongside source links or search result recommendations only, depending on whether the system had sufficient content to generate a response. 

We conducted a three-phase sequential design. First, we analyzed \textsc{Steve} and \textsc{People Tool} search logs to identify behavioral patterns across query formulation, result availability, click-through behavior, and GenAI answer exposure. Based on these patterns, we conducted an exploratory survey with 25 employees across six countries to assess usage, usefulness, usability, trust, system preference, and employee reactions when the answer generated by GenAI was insufficient. Afterward, we conducted semi-structured interviews with 10 employees from four countries to explore how employees perceive the two systems and how personal, organizational, and technological factors affect adoption paths. Since each method served a different purpose, the three data sources cover different country contexts and are treated as complementary rather than directly comparable country-level samples. This study received ethical approval from our institution, and all data were handled in accordance with GDPR requirements.

\subsection{Search Log Analysis}
Search logs are system-level logs that capture information about employees' search activities. Each search log recorded details of queries entered by employees, the systems in which queries were searched, the search timing, country context, and system feedback, such as whether the system returned results or displayed a GenAI response. Search log analysis served as the pre-study phase of our sequential mixed-methods design. Its purpose was exploratory: to surface behavioral patterns in how employees queried each system, with findings directly shaping the design of the survey instrument and the interview protocol. 

We obtained search logs from both systems covering six countries: Egypt, the Netherlands, the Philippines, Pakistan, Qatar, and Saudi Arabia. \textsc{Steve} logs spanned approximately one year (September~2024--August~2025). \textsc{People Tool} logs covered a two-week window (July~24--August~7,~2025) due to a hard constraint imposed by the system provider's data retention policy. For cross-system volume comparisons, we restricted both datasets to the common observation window (July~24--August~2,~2025) to avoid conflating differences in system use with differences in observation period; the longer \textsc{Steve} baseline was retained separately to characterize historical query patterns and queries that returned ``no result'' prior to the introduction of \textsc{People Tool} (Appendix~\ref{app:log-analysis}).

Non-English queries were translated into English by a native speaker to allow topic-level comparisons across countries. The analysis focused on query volume and distribution across the two systems, query phrasing and length, result availability, GenAI response exposure rates, and topic overlap between the two systems. The logs record what employees searched and the results obtained from each system; they do not provide information about why employees chose one system over the other, nor about how they perceived those results; those questions were addressed in the survey and interviews.

\subsection{Survey}
Building on the initial analysis of the search log, we conducted an exploratory survey of employees who could use the \textsc{People Tool} and its built-in GenAI search functionality. Participants were recruited from six countries: Germany, the US, India, Austria, Switzerland, and France. This survey was designed to accomplish two tasks: collecting descriptive data on employees' experiences with \textsc{Steve} and \textsc{People Tool}, and refining the interview guide for the upcoming interview phase.

The survey collected contextual information, including country of work, job level, and tenure. Employees were required to report on their usage frequency for each system, their top HR topics searched for, and their ratings on the systems' ease of use, helpfulness, and trustworthiness. Specific attention was paid to the use of the GenAI search within the \textsc{People Tool}, where respondents answered on how often they received an answer by the GenAI function, how helpful the answers were, how often they clicked on sources presented after answering, and what action they would take in case the answer was insufficient. The respondents were free to report any problems they faced, their preferences between the two systems, topics that did not yield useful results, and areas for improvement. Finally, a call was made to those willing to participate in the interviews.

The survey was answered by 25 employees. For the closed-ended questions, we conducted descriptive statistics using counts and percentages. For the open-ended questions, we analyzed responses to identify recurring concerns that might reappear in interviews. The survey results cannot be considered statistically representative due to their limited sample size and unequal representation. Instead, they serve as descriptive evidence for the interviews and help connect behavior identified in the logs with the process discussed in the interviews (Appendix~\ref{app:survey-analysis}, Table~\ref{tab:key-survey-descriptives}, Limitations Section~\ref{sec:lim}).

\begin{table}[h]
\centering
\caption{Key descriptive survey results used to contextualize the findings.}
\label{tab:key-survey-descriptives}
\footnotesize
\setlength{\tabcolsep}{3pt}
\renewcommand{\arraystretch}{1.08}
\begin{tabularx}{\columnwidth}{@{}Xcc@{}}
\toprule
\textbf{Measure} & \textbf{\textsc{People Tool}} & \textbf{\textsc{Steve}} \\
\midrule
Daily or weekly use        & 19/25 (76\%) & 14/25 (56\%) \\
Preferred system           & 14/25 (56\%) & 3/25 (12\%)  \\
Very/extremely helpful     & 16/25 (64\%) & 8/25 (32\%)  \\
Easy/very easy to use      & 16/25 (64\%) & 16/25 (64\%) \\
Mean helpfulness           & 3.76         & 3.12         \\
Mean trust                 & 3.92         & 3.72         \\
\midrule
\multicolumn{3}{@{}l}{\textit{GenAI-specific items}} \\
\midrule
\multicolumn{2}{@{}l}{Never/rarely clicked sources in GenAI answers}
  & 13/25 (52\%)  \\
\multicolumn{2}{@{}l}{Would rephrase when GenAI answer was poor}
  & 19/25 (76\%)  \\
\bottomrule
\end{tabularx}

\begin{minipage}{\columnwidth}
\footnotesize
\emph{Note.} Survey data are descriptive, not statistically representative. Preference responses were mixed or unclear for 7/25 (28\%). Helpfulness and trust means are on 5-point scales (see Appendix~\ref{app:survey-analysis}).
\end{minipage}
\end{table}

\subsection{Interview Study}
Semi-structured interviews were the primary source of interpretive data in this research. While search logs and surveys provided information on behavior and contextual descriptions, interviews helped examine how people interpreted \textsc{Steve}, \textsc{People Tool}, and the GenAI search feature in practice. Interviewees were selected among those who agreed to participate in the interviews in addition to the survey, as well as from employees involved in everyday use of \textsc{People Tool}.

We conducted ten semi-structured interviews using an online videoconferencing tool. Each interview was recorded, lasted between 20 and 35 minutes, and was conducted in English or Spanish, at the participant's preference. At the beginning of each interview, participants were reminded of the purpose of the study, informed that participation was voluntary, and asked to consent to the recording for transcription and analysis.

The interview guide was organized around six key themes, which involved participants’ background and familiarity with digital technologies; participant experience and comparison of \textsc{Steve} and \textsc{People Tool}; perceived usefulness and limitations of GenAI on Search; trust, source-checking, and information behavior; adoption dynamics and organizational context; and reflections on future improvement.

Interview participants represented two broad perspectives. The first group comprised employees involved in strategy, rollout, knowledge management, system ownership, testing, or support for \textsc{People Tool}; however, these participants also used the system in their own work, and could therefore speak both as practitioners with implementation knowledge and as users seeking HR-related information. The second group comprised employees whose engagement with the systems was primarily as users seeking HR-related information. We use this distinction to contextualize participants' accounts, but we do not conduct a systematic comparative analysis between these groups. Participants varied in tenure, ranging from less than 1 year to more than 15 years, and were distributed across four countries. Table~\ref{tab:participants} provides an overview of the interview participants.

We generated the interview transcriptions automatically and manually checked them for accuracy. We anonymized them by removing all personal identifiers and assigned new participant IDs ranging from P1 to P10. Minor corrections were made to enhance readability without altering the content bearing the message. To analyze the interview data, we used reflexive thematic analysis following Braun and Clarke~\citep{braun2006thematic}. We first read the transcripts repeatedly to familiarize ourselves with the data, then generated descriptive codes capturing participants' accounts of system comparison, perceived usefulness, ease of use, trust, source checking, fallback practices, AI-search literacy, training and communication, organizational support, content quality, work context, language, and attitudes or fears towards AI. These initial codes were then iteratively grouped into more focused sub-themes. 

This analysis was conducted over multiple rounds of discussions among four researchers. Instead of considering coding as a single classification process, the discussions helped us compare understandings, merge redundant codes, push against poorly formed themes, and link sub-codes to broader analytical themes that relate to the research question. The outcome of this discussion led us to the final themes presented in the Findings section: selective use across parallel systems, situated fit across workers and contexts, learning a new GenAI search logic, reliance on checking and fallback, and knowledge conditions.


\begin{table}[h]
\centering
\caption{Overview of interview participants ($n=10$).}
\label{tab:participants}
\footnotesize
\setlength{\tabcolsep}{4pt}
\renewcommand{\arraystretch}{1.12}

\begin{tabularx}{\columnwidth}{@{}l Y c l@{}}
\toprule
\textbf{ID} & \textbf{Generalized role context} & \textbf{Tenure (yrs)} & \textbf{Country} \\
\midrule

\multicolumn{4}{@{}l}{\textit{Project team participants}} \\
\addlinespace[1pt]
P1  & HR IT / digital transformation; application ownership & 1--5  & Germany \\
P2  & Transition project member & 5--15 & Germany \\
P3  & HR IT intern; content migration and AI testing & 1--5  & Germany \\
P4  & Product ownership for GenAI and knowledge management & 15+   & Germany \\
P7  & HR support leadership & 5--15 & Spain \\
P8  & Regional HR IT coordination & 5--15 & USA \\

\addlinespace[3pt]
\midrule
\multicolumn{4}{@{}l}{\textit{End-user participants}} \\
\addlinespace[1pt]
P5  & Digitalization-related role in adjacent ServiceNow environment & 5--15 & USA \\
P6  & Compensation and HR tooling role & 5--15 & USA \\
P9  & HR systems project management and cloud migration & 1--5  & Spain \\
P10 & Early-career P\&O systems professional & $<$1 & Norway \\

\bottomrule
\end{tabularx}

\vspace{1mm}
\begin{minipage}{\columnwidth}
\footnotesize

\emph{Note.} Role descriptions are intentionally generalized to preserve participant anonymity.
\end{minipage}
\end{table}
\section{Findings}

\begin{figure*}[t]
    \centering
    \includegraphics[width=\linewidth]{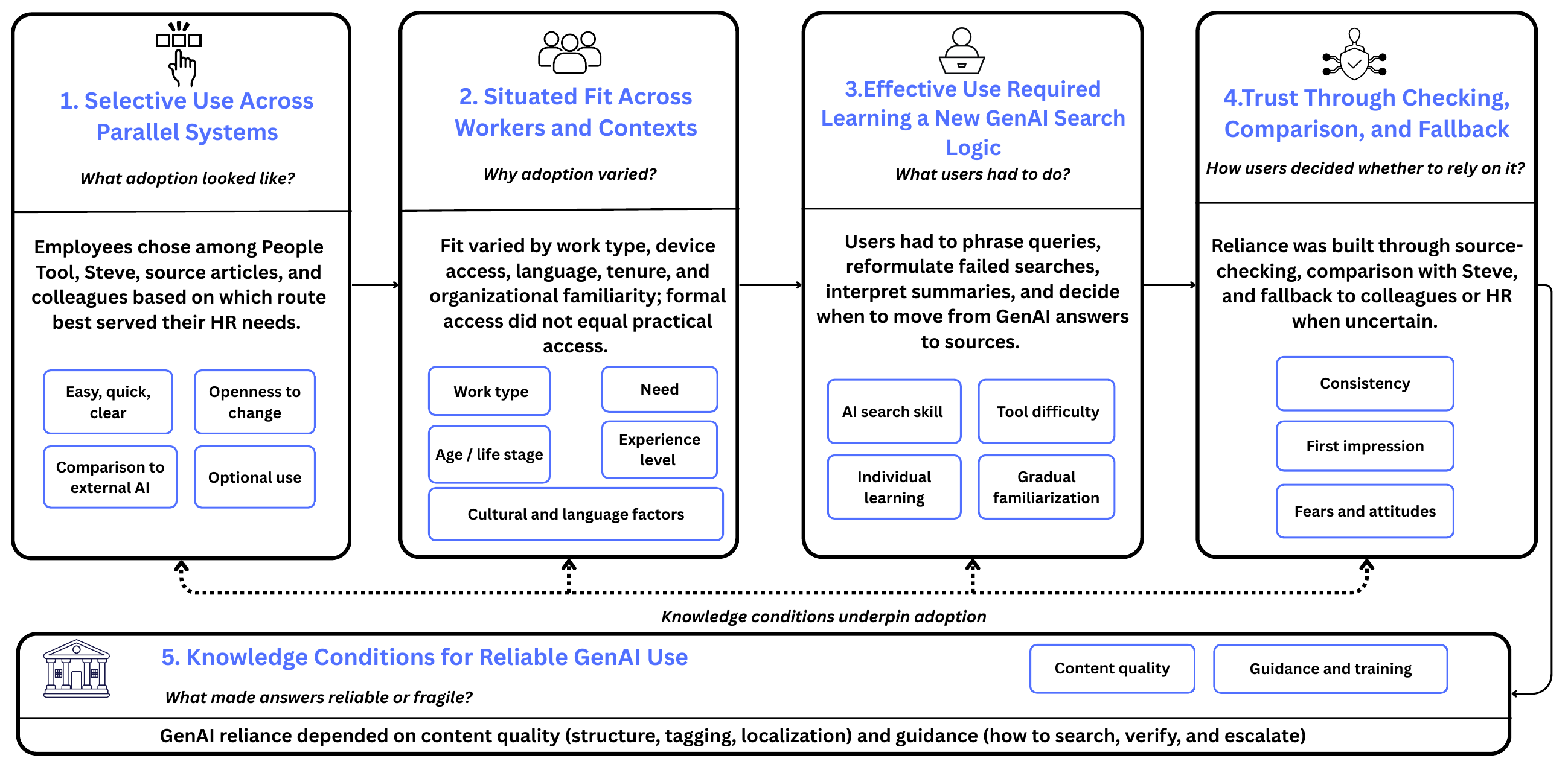}
    \caption{Adoption unfolded through selective use across parallel systems, varied by situational fit, required learning a new GenAI search logic, and depended on trust-building through checking, comparison, and fallback. Knowledge conditions underpinned the whole process.}
    \label{fig:adoption-framework}
\end{figure*}

By analyzing interviews, survey responses, and the search logs of both systems, we observed that adoption was not a linear process of migration from legacy search to GenAI-enhanced search. As both systems, \textsc{People Tool} and \textsc{Steve}, were available, employees could switch between them, compare their outcomes, verify the responses, and return to alternative routes when necessary. Adoption was shaped by the interaction between employees, the system, and the organization.

In this section, we first show that adoption involved selective use across parallel systems rather than a complete migration to the new system. Then, we examine why this use varied across employees and contexts. Next, we argue that effective use requires employees to learn a new GenAI search logic. This, in turn, raises questions of trust, as employees had to determine when an answer was reliable enough to act upon. Finally, we show that these everyday adoption practices were shaped by organizational knowledge conditions.

\subsection{Selective Use Across Parallel Systems}

Since \textsc{Steve} continued to function after the transition, employees were not required to rely on a single route for HR information at all times. Instead, they chose among the available routes depending on the situation. For instance, employees would use \textsc{People Tool} if it provided speed, integration, and clearer direction. They would use \textsc{Steve} if the search process were familiar, the results were predictable and article-based, or if it were a route they already knew how to navigate. Source articles provided precise information when needed, and employees reached out to colleagues, managers, or the HR department for personal or complex answers. Thus, adoption was not a single decision point, but an ongoing practice.

Survey responses reflected this parallel-use pattern. Indeed, 76\% of the respondents (19/ 25) used \textsc{People Tool} either daily or weekly. However, \textsc{Steve} was not discontinued as 56\% of respondents (14/25) used \textsc{Steve}  daily or weekly. Further, regarding preferences, 56\% of respondents preferred \textsc{People Tool}, whereas 12\% preferred \textsc{Steve}. It appears that although \textsc{People Tool} was widely used, the legacy system was still used as part of the routine HR information-seeking workflow.

Interviews also elucidate the underlying logic by showing that switching to a new system was worthwhile. Employees were not asking only whether \textsc{People Tool} was new or AI-powered, they were interested in its advantages relative to familiar information-search methods. One participant explained his concerns as follows:

\begin{quote}
\textit{``Oh man, now there's another HR system (\textsc{People Tool}) I have to use. You know, I'm comfortable with this one. Why do we have to change again?''} (P5)
\end{quote}

However, this response cannot be regarded as a sign of general resistance to AI. Learning a new route takes time and effort that employees simply do not have to devote to a task of lower priority than their main work. Since a familiar route could be accessed with a single click, the new route needed to provide sufficient value. If \textsc{People Tool} offered clear answers and useful information, employees could find a valid reason to use it; but when it added more navigation and cognitive burden without a clear benefit, familiar routes remained the easier choice.

Therefore, the simultaneous availability of \textsc{People Tool} and \textsc{Steve} meant that adoption was not a single migration decision from the legacy to the new GenAI system, but an ongoing practice of choosing among routes. This raises a question: \textit{what made \textsc{People Tool} the preferred route for some employees but not others?}


\subsection{Situated Fit Across Workers and Contexts}

The reasons behind selective adoption were rooted in situational fit. We define situational fit as the degree to which the tool's offerings match users' actual work conditions, needs, and contexts. \textsc{People Tool} was designed for laptop and computer access, assumed organizational familiarity, and prioritized some languages over others, resulting in a fit that varied considerably across the workforce.

A recurring concern throughout the interviews was that \textsc{People Tool} was better suited to office-based employees who regularly had access to laptops, were accustomed to the company's internal knowledge system, and could frame their HR inquiries in organizational terminology. For shift-based, factory workers, and shop-floor employees, this was not the case. Participants suggested that these worker groups relied on their managers, team assistants, local HR staff, or informal channels. For instance, one participant noted that blue-collar workers may not be familiar with organizational HR terminology, making it difficult to even formulate a question \textit{``when you're a blue-collar, you're not as exposed to that side of things and so even knowing what to ask''} (P6). Another participant suggested that \textsc{People Tool} was \textit{``not ideal''} for factory workers (P1). These responses suggest that although the system was technically available to all employees, its design for office-based, digitally familiar, and organizationally embedded users could lead to exclusion, with employees with diverse work conditions at a disadvantage in using the tool.

Thus, making \textsc{People Tool} formally available to employees did not mean that all employees could use it equally in practice. Some employees lacked regular access to a suitable device \textit{``workers who go on site to work...they sometimes don't even have a laptop''} (P5); others did not know how to phrase HR questions in organizational terminology (P6), and others continued to rely on managers, local HR staff, or colleagues as their main route to information. These employees could disappear from system usage logs or aggregate adoption figures, even if the tool was technically available to them. As a result, aggregate adoption figures therefore overstate inclusion, as a high query volume from office-based employees masks the absence of shift-based, blue-collar, and multilingual workers who never meaningfully engaged with the system.

Tenure and life stage played an important role in shaping situational fit as well.
Because new recruits were still learning the organization's internal processes and the location of HR documentation, they could gain the most from \textsc{People Tool}. On the other hand, tenured employees used the \textsc{People Tool} less, not because they resisted it, but because their information uncertainty had declined, and they already knew the company's policies, the personnel responsible for addressing HR concerns, and the routine mechanisms to resolve issues. As one participant explained: 
\begin{quote}
\textit{``The older you are, the less changes you have in life. I would say the kids are gone, house is built, car is there. What shall you ask HR? Maybe the younger people have more interactions with HR because they are still building their lives.''}
(P4)

\end{quote}

The GenAI tool's situated fit with employees was also evident from the logs. First, search behavior was distinct by country. For instance, in the Netherlands, users searched for policy and process-related topics, in Saudi Arabia and Qatar, searches focused on documents and certificates, in Egypt, on benefits and insurance, and in Pakistan, on travel claims. System-level indicators were distinct as well, for instance, in Pakistan, click-through was the lowest across all countries at 18.5\%, suggesting employees were less likely to move from search results to further interaction with the system. 
These patterns show that a global HR search system encountered diverse local task environments, with employees searching for HR artifacts using distinct terminologies and addressing diverse requirements. 



Language was another factor that impacted adoption. If a system failed to accommodate the user's language or did not reflect local HR terminologies, its effectiveness suffered despite its technical capabilities. In multinational rollouts, language support is typically added incrementally, with non-English and non-European languages often deprioritized or absent altogether; one participant emphasized that \textit{``if you don't have the languages, then people sometimes don't use it''} (P4). Together, the interview and log data suggest that a one-size-fits-all design for GenAI is ineffective in multinational HR knowledge work. The same system provided convenience to some users while creating friction or exclusion for others.

Thus, situational fit shaped the selective use of \textsc{People Tool}. When the tool's offerings aligned with users' device access, language, work environment, life stage, and HR information needs, it gained greater traction. Otherwise, employees relied on legacy systems or local sources that they were already familiar with. This raises the next question, \textit{even when the tool aligned with a user's context, what did it take for employees to effectively use GenAI HR search?}

\subsection{Effective Use Required Learning a New GenAI Search Logic}

Access to \textsc{People Tool} did not automatically translate into effective use. Even when system fit a user's context, employees still had to learn a new way of searching that differed fundamentally from \textsc{Steve}. This involved phrasing questions, adding context, rephrasing failed queries, interpreting summaries, and knowing when to move from a GenAI response to source articles or human support. Thus, adoption was based not only on what the system offered but also on whether employees could translate their HR needs into questions the system could answer; and as the survey data show, these two dimensions usefulness and ease did not always align.

This distinction helps explain why usefulness and ease did not necessarily occur in parallel. The survey showed that \textsc{People Tool} was rated as ``very'' or ``extremely'' useful by 64\% of respondents (16/25), compared with 32\% (8/25) for \textsc{Steve}. However, both systems were rated as ``easy'' or ``very easy'' by the same proportion of respondents, 64\% (16/25). In other words, employees recognized the added value of \textsc{People Tool} GenAI search, but realized that value required learning a new interaction logic, including how to ask, interpret, and adjust their searches. 

The survey further illustrates this interaction effort. When GenAI failed to provide a satisfactory answer, 76\% of the respondents (19/25) reported they would rephrase their question. This demonstrates that users viewed poor answers as something that could be improved by asking differently, rather than a system failure. By doing so, employees participated in creating the response, making the answer dependent not only on the system itself, but also on employees' knowledge of what to ask and how. The remaining 24\% did not rephrase, suggesting that some employees disengaged entirely when the system failed to provide a useful answer. 

The logs show that this new search logic had not yet fully replaced older keyword-search habits. We measured query length in terms of tokens, defined as units similar to words in a search request. The average query length increased only slightly, from 1.87 tokens in \textsc{Steve} to 2.11 tokens in \textsc{People Tool}. Similarly, the share of short one- to two-token queries stayed practically unchanged, 83.4\% in \textsc{Steve} and 83.3\% in \textsc{People Tool}. This suggests that employees continued to use traditional keyword-based search practices even in the GenAI-enabled environment, although short keyword queries may also have remained effective given the system's retrieval-augmented architecture. The interviews explain this, as a participant described the change as a shift in search logic:

\begin{quote}

\textit{``Once they [the user] have understood that the logic changes, that the way of searching and finding is different, I think there is greater acceptance''} (P7)

\end{quote}

Effective use, therefore, required more than familiarity with the system; employees had to develop a new form of search competence, including learning not only where to search, but how to search with AI, how to recover when the answer was poor, and how to judge whether an answer was enough for their HR question. The survey evidence supports this; while 64\% of respondents rated \textsc{People Tool} as useful, the same proportion rated both systems as equally easy to use, suggesting that perceived value and interaction competence are distinct dimensions of adoption. This competence was not about learning where to search, but about developing an ongoing practice of prompting, evaluating, and deciding. This leads to the next issue: even when employees learned how to ask, \textit{how employees decided whether a GenAI answer was reliable enough to use?}

\subsection{Users Built Trust by Checking, Comparing, and Fallback}

Receiving GenAI answer was not sufficient for adoption. Employees also had to decide whether that answer was accurate enough to act on. In HR, acting on incorrect or partial information can have serious implications for salary, leave, and personal circumstances. Therefore, adoption depended on employees' ability to move from obtaining an answer from the GenAI tool to deciding whether to rely on it. This reliance was not based on blind acceptance but was calibrated through checks, comparisons, and fallback.

The survey suggests a gap between usefulness and trust. Mean helpfulness was higher for \textsc{People Tool} than \textsc{Steve}: 3.76 compared with 3.12. By contrast, the difference in mean trust was more modest, 3.92 compared with 3.72. This suggests that employees may perceive greater value in \textsc{People Tool} while remaining cautious about whether its answers are complete, current, and applicable to their situation. This caution also appeared in the interviews as one participant stated:

\begin{quote}

\textit{``I will not sell my house because of an AI answer. There’s always a double net. There’s always checks and balances, talking to managers, talking to colleagues, talking to my wife, consulting someone, friends who made this already. So it’s not just the answer in [People Tool] leading to the decision. It’s always a mix of information and then you come to your own conclusion.''} (P4)

\end{quote}

Source-checking was one way of managing this uncertainty, although it was not used consistently. In the survey, 52\% of respondents (13/25) reported that they never or rarely clicked the sources shown in GenAI-generated answers, while 48\% (12/25) clicked them sometimes or often. This difference should not be viewed simply as a split between trusting and distrusting users; sometimes, employees did not click sources because the GenAI answer was sufficient for their needs and required no further verification. Others coped with this uncertainty in different ways, such as inspecting sources, relying on plausibility, restructuring their question, comparing the answer with what is known from other systems, and seeking help from others.

At the same time, trust was not always actively calibrated. Some employees appeared to lend credibility to \textsc{People Tool} because it was an internal company system. As one participant explained: \textit{``I think people innately trust things internally in [the Company] that it's going to give them the correct answer, and I think people are trusting AI too much.''} (P6). Yet, the organization placed responsibility for interpreting outputs back onto employees as one participant pointed out: \textit{``We always tell the users, `You can use AI, but the risk is on you.' How can I force them to make use if they know that they have the responsibility?''} (P4). Employees sometimes trust \textsc{People Tool} GenAI responses not because they have reviewed their source or assessed their completeness, but because they were delivered through an approved organizational platform; this, in turn, creates the risk that fluent but incorrect answers would be accepted without sufficient verification.

The interviews also identified the continued need for a fallback.  A GenAI summary could be useful for quick orientation, particularly when the question was general or low-risk. But when questions are specific, complex, or high-risk, employees find greater satisfaction in seeking reassurance from source documents, \textsc{Steve}, peers, supervisors, or HR. Hence, trust was distributed across the answer itself and the wider set of information routes surrounding it.

Trust was also shaped by prior experiences and first impressions. Some employees approached GenAI with curiosity, while others began with skepticism, privacy concerns, or worries about unreliable HR guidance. These starting points did not determine adoption on their own, but they influenced how employees interpreted their initial interactions with the system. Repeated use could revise initial expectations, but only when the system behaved consistently enough to support confidence. When GenAI answers were wrong or irrelevant, employees would disengage quickly. When answers appeared plausible but remained uncertain, employees had to decide whether to verify, rephrase, or ask someone else.

In general, reliance on \textsc{People Tool's} GenAI search was built through repeated use rather than assumed a priori. Employees delegated some information work to \textsc{People Tool}, but also checked, compared, or returned to other routes when uncertain. This raises another question: \textit{what made GenAI's answers reliable or unreliable in the first place?}

\subsection{Knowledge as a Condition for Reliable GenAI Use}
The previous analysis showed that employees needed mechanisms to validate, compare, and trace answers back to different sources when GenAI-produced answers are ambiguous. Thus, there was an even deeper issue about the trustworthiness of \textsc{People Tool} that was not dependent on the GenAI feature alone. Employees accessed HR information through GenAI, which in turn depended on the quality of the underlying HR articles, metadata, tagging, taxonomy, localization, training, communication, and escalation routes. When these were well-structured, this allowed the generation of useful guidance via the GenAI process; however, uncoordinated input data, outdated articles, poor tagging, and vague explanations made it difficult for the GenAI tool to compensate for these issues.

This is supported by survey data, which shows that while the \textsc{People Tool}  proved to be more effective than \textsc{Steve} (mean helpfulness of the tool has risen from 3.12 to 3.76), trust increased only moderately (up from 3.72 to 3.92). 
In other words, \textsc{People Tool} made HR information easier to find, but employees remained uncertain about whether the answers were complete, current, and applicable to their specific situation. Employee responses focused on the quality of information architecture more than on the AI technology itself, as one employee noted:
 
\begin{quote}
\textit{``In \textsc{People Tool} the information is not that user-friendly and as each item might be under a different article, it's hard to find what you want.''} (Survey respondent)
\end{quote}

Another employee linked poor search results to weak tagging, \textit{``Not every document or article has the tags that I would search for and therefore they are not shown in the search results''} (Survey respondent). These responses illustrate why GenAI reliability cannot be separated from content quality. Employees sometimes attributed poor answers to the AI when the underlying problem was poorly organized, outdated, or localized HR content.


Training and guidance were also part of these organizational conditions, not just rollout activities. Employees needed to know more than that \textsc{People Tool} existed; they needed to understand what had changed, how to search, when GenAI answers would appear, when to check sources, and when to escalate to human support. Interviewees suggested that support was most useful when provided close to the moment of use, through examples, short guidance, Q\&A sessions, office hours, prompts, and clear contact routes. One-off pre-launch communication was less effective because it could be forgotten before employees actually needed the tool.

Therefore, content quality, guidance, and training were not background conditions for GenAI adoption; they determined whether the answers employees received were reliable enough to act on. \textsc{People Tool} could make HR search faster and more convenient, but only when the underlying content was structured, maintained, localized, clearly communicated, and employees were supported at the point of use.


\section{Discussion}
\subsection{Who Benefits from Workplace GenAI?}
Recent research on AI fairness and sociotechnical design argues that potential harms arising from AI cannot be understood only by examining model behavior; the social and organizational context in which these systems are deployed must also be considered \citep{selbst2019fairness,costanzachock2020design}. Similarly, work in workplace HCI has shown how algorithms affect workers' experiences within organizations and the management settings and infrastructure of workplaces \citep{cheon2025warehouse}.  Research on digital inclusion has also highlighted the mistake of considering technological access alone is sufficient for understanding inclusion and emphasized the roles of skills, social context, conditions of use, and the ability to translate technologies into concrete gains \citep{warschauer2003technology,vandijk2005deepening,helsper2021digital}. Our findings extend this argument to workplace GenAI deployment: making a GenAI system available across an organization does not mean that all employees are equally positioned to benefit from it.

In our study, formal availability of the GenAI HR system did not translate equally into useful HR support. As highlighted in Section 4.2, employees’ ability to take advantage of the system depended on its alignment with work processes, linguistic preferences, HR-related information requirements, and access to support channels. This raises an equity issue that remains challenging to operationalize. \textit{When the system is designed for office-based, digitally confident, and English-speaking users, how will organizations recognize which workers are at a disadvantage?} This question is difficult to answer through aggregate usage metrics alone, because those who face the greatest friction may be characterized by their absence, inactivity, and infrequent use. A system may therefore appear broadly adopted while still providing more reliable and actionable support to some groups and exclude the others \citep{selbst2019fairness,HumlumVestergaard2025}. The relevant question is not only who can access the system, but who can obtain useful, trustworthy, and actionable support from it in everyday work.

\textbf{Design recommendations}. Organizations should evaluate their GenAI deployment in the workplace by considering role-sensitive and context-sensitive benefit measures. This includes examining which groups are missing from usage logs, which users receive search results but not GenAI answers, which queries fail or require repeated reformulation, who continues to rely on local intermediaries, and which groups do not receive useful or actionable support. Then, the less visible employee segment should be included continuously in testing, feedback, and redesign, rather than being treated as edge cases after deployment.

\subsection{Who Carries the Risk of Getting It Wrong?}

Previous studies on human-AI interaction show that effective use of AI involves calibrated reliance, avoiding both over-reliance on incorrect outputs and under-reliance on useful ones \citep{lee2004trust,bansal2021does}. The factors that impact trust are not limited to system efficiency or performance but also include information communicated about the AI system, such as explanations, source visibility, and institutional credibility \citep{liao2022match}. With GenAI specifically, users face additional metacognitive demands, including prompting, evaluating outputs, and deciding whether an answer is reliable enough to act on \citep{tankelevitch2024metacognitive}. However, in reality, as illustrated by our findings (Section~4.4), reliance extended beyond the user-system interaction; we observed that employees were moving between the GenAI system \textsc{People Tool}, legacy systems \textsc{Steve}, documents, coworkers, and HR  personnel, depending on task priority and uncertainty. This raises a question worth discussing: \textit{ When an employee navigates this network for HR information and still acts on incorrect information, who is responsible? The system designer, the organization that deployed it, or the employee who used it?}

This question is consequential in HR contexts, where employees may act on AI-generated information about high-stakes decisions such as pay, leave, benefits, or employment conditions \citep{budhwar2023hrmgenai, dasaklis2025llmHRM}.  When the AI-generated answers appear fluent and delivered through an approved system by the organization, institutional trust can replace the careful verification by the employee \citep{knowles2021sanction, liao2022match}.  Without clear information on source legitimacy, established escalation procedures, and robust accountability structures within the organization, responsibility for accuracy rests with employees rather than with the organization that deployed the system.

\textbf{Design recommendations.}  At a system design level, workplace GenAI systems should make provenance, uncertainty, and fallbacks apparent as default features. Source URLs need to be seen as an accountability architecture, not simply a UI-based feature. Also,  the GenAI answers themselves must provide information about their origin, verifiability, and whether human intervention is recommended.  At the organizational level, it is important to maintain existing human channels for assistance where possible, rather than replacing them with AI.  It is also critical for organizations to establish clear lines of accountability when AI systems are involved in decision-making in high-risk areas, so that the extent to which the system can and cannot be relied upon is clear, and interpretive risk does not fall on individual workers by default.

\subsection{Knowledge Readiness as AI Infrastructure} 

As stated in (Section~4.5), participants did not view \textsc{People Tool} as a stand-alone AI model. Users interacted with \textsc{People Tool} through HR artifacts, including articles, metadata, tagging, taxonomy, localization, source references, training guides, communication, and escalations. When these elements were well structured, GenAI could make HR information faster and easier to access. On the contrary, when some of these artifacts were absent, out-of-date, poorly labeled, or difficult to navigate, the GenAI system would generate incomplete answers or answers that did not address employees' needs. 

This means that knowledge management becomes a critical part of accountability for workplace GenAI implementations \citep{alavi2024genaiKM}. In high-stakes domains, content quality is more than a back-office implementation issue; it determines whether the information summarized by AI systems is complete, accurate, up-to-date, and reliable for action \citep{tankelevitch2024metacognitive}. Similarly, training and communications are not limited to mere rollout considerations. Employees need to be able to frame their questions, interpret the results generated, check their sources, and raise uncertainties. Without such considerations, companies may place AI-related uncertainties on individuals' shoulders. 

\textbf{Design recommendations.} Knowledge infrastructure should be considered part of the core AI infrastructure. This includes keeping HR content up to date, assigning ownership of articles, improving tagging and taxonomies, localizing terminology, auditing unsuccessful queries, and establishing feedback loops between poor AI answers and knowledge owners. Training must be continuous and timely with respect to its use. This involves using brief instructions, examples, rewording help, Q\&A, and readily available help lines. In corporate GenAI applications, responsible deployment requires preparing the organization around the AI, not just for the AI itself.

\section{Limitations} 
\label{sec:lim}

This study has several limitations. First, it is based on a single organizational case. This limits statistical generalizability, but the case is valuable because it captures GenAI adoption during a live workplace transition, where employees could compare the new system with an existing legacy system. Second, the number of survey respondents (n = 25) was relatively small and unevenly distributed across nationality, job roles, and work experience within the company. For that reason, the survey data were treated as descriptive only and triangulated with data from interviews and search logs to enable interpretation. Third, the logs captured different usage periods for both systems. While \textsc{Steve}'s logs capture a longer span of searches, \textsc{People Tool}  logs had to be confined due to ServiceNow's retention policies. To address this limitation, we restricted cross-system comparisons to the same observation window and used the longer \textsc{Steve} logs only as historical context. Finally, some worker groups central to situational fit, such as shop-floor, shift-based, or less digitally embedded employees, were not systematically sampled as direct participants.  Some interviewees held roles that exposed them to these work contexts, so their accounts provide informed yet indirect evidence of uneven fit. Future studies would do well to include such groups directly to study whether workplace GenAI systems reproduce or reduce inequalities across different labor conditions.

\section{Conclusion}
In this study, we examine how employees across a multinational workforce adopt, resist, and selectively use a GenAI-enhanced HR knowledge system during a live organizational transition from a legacy HR knowledge system to a new GenAI-enhanced system. Through our analysis of both systems' search logs, a survey, and ten semi-structured interviews, we show that several factors shaped the adoption process, including situational fit, search literacy, trust calibration, and deployment conditions, such as content quality, user guidance, and accountability structures. Drawing on these findings, we argue that equitable workplace GenAI deployment cannot be assessed solely by access to the system. Our results raise questions about how organizations should define adoption success, who is responsible for wrong AI-generated HR information, and what conditions are necessary for GenAI to be deployed reliably and equitably in high-stakes domains.
\newpage

\clearpage

\section{Researcher Positionality Statement}
\textit{Our research team combines expertise in human-computer interaction (HCI), responsible and sociotechnical AI, organizational studies, computational social science, and human resource management, spanning both the Global North and the South. The team members include PhD and MSc students, faculty, an industry researcher, and an HR practitioner from five countries. One of our team members has a prior professional relationship with the organization, which facilitated data access; interpretive decisions were made collectively and without organizational involvement. Disagreements during the thematic analysis stage were resolved through iterative discussion among all researchers.}

\section{Ethical Statement}
\textit{This study received ethical approval from our institution prior to data collection, and all data were handled in accordance with GDPR requirements. All participation in the survey and interviews was voluntary with no financial incentive, and informed consent was obtained from all interview participants before recording. All identifying information was removed from transcripts and participants were assigned anonymous IDs. Search log data were provided by the organization and contains no personally identifiable information. The organization reviewed and approved the data sharing arrangements prior to analysis.}
\bibliography{ref_used}

\newpage

\clearpage
\appendix

\section{Descriptive Search Log Analysis}
\label{app:log-analysis}

Descriptive statistics on the results of the search log analysis are provided in this appendix. The search log analysis helped formulate the survey and interview guide by examining how employees performed searches within the old system, \textsc{Steve}, and the new GenAI-assisted system, \textsc{People Tool}. It is essential to understand that these results reflect search behaviors, access to search results, clicking patterns, and exposure to GenAI responses, but not measures of trust, satisfaction, motivation, or adoption intention. The two systems cover different spans of time, which means that any comparisons between them are purely descriptive and directional.

\subsection{Log Dataset Overview}

Table~\ref{tab:appendix-log-overview} summarizes the two log datasets used in the analysis.  While the legacy system holds a more extensive history, the new system holds only a limited history due to platform retention restrictions. 

\begin{table}[h]
\centering
\caption{Overview of log datasets.}
\label{tab:appendix-log-overview}
\scriptsize
\begin{tabularx}{\columnwidth}{p{3.99 cm}X}
\toprule
\textbf{Item} & \textbf{Description} \\
\midrule
Legacy system & \textsc{Steve} \\
Newer system & \textsc{People Tool} \\
\textsc{Steve} window & Sept.~2024--Aug.~2025 \\
\textsc{People Tool} window & July~24--Aug.~7, 2025 \\
Contexts & EG, NL, PH, PK, QA, SA \\
\textsc{Steve} size & 17,040 query rows; 19,562 weighted searches \\
\textsc{People Tool} size & 1,226 query rows \\
\textsc{People Tool} GenAI-displayed subset & 343 query rows \\
\textsc{Steve} fields & Search term, country, language, date, search count, result count \\
\textsc{People Tool} fields & Query, country, language, timestamp, has-results flag, click rank, GenAI answer displayed flag \\
\bottomrule
\end{tabularx}
\vspace{1mm}
{\raggedright\scriptsize \textit{Note:} The GenAI answer displayed flag means that a GenAI-generated answer appeared in the interface. It does not indicate that the user read, trusted, or acted on the answer. Country codes: EG~=~Egypt, NL~=~Netherlands, PH~=~Philippines, PK~=~Pakistan, QA~=~Qatar, SA~=~Saudi Arabia\par}
\end{table}

\subsection{Query Behavior Across Systems}

\begin{figure}[h]
\centering
\begin{tikzpicture}
\begin{axis}[
    ybar,
    width=\columnwidth,
    height=4.5cm,
    ymin=0,
    ymax=2.5,
    ylabel={Avg. tokens},
    symbolic x coords={Steve,People Tool},
    xtick=data,
    nodes near coords,
    nodes near coords align={vertical},
    bar width=18pt,
    enlarge x limits=.35
]
\addplot coordinates {(Steve,1.87) (People Tool,2.11)};
\end{axis}
\end{tikzpicture}
\caption{Average query length in \textsc{Steve} and \textsc{People Tool}.}
\label{fig:appendix-query-length}
\end{figure}
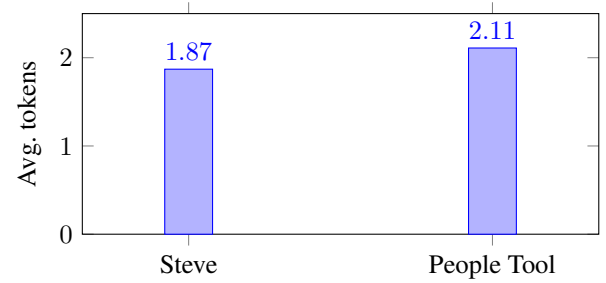

\begin{figure}[h]
\centering
\begin{tikzpicture}
\begin{axis}[
    ybar,
    width=\columnwidth,
    height=4.5cm,
    ymin=0,
    ymax=100,
    ylabel={Queries (\%)},
    symbolic x coords={Steve,People Tool},
    xtick=data,
    nodes near coords,
    nodes near coords align={vertical},
    bar width=18pt,
    enlarge x limits=.35
]
\addplot coordinates {(Steve,83.4) (People Tool,83.3)};
\end{axis}
\end{tikzpicture}
\caption{Share of one- to two-token queries in \textsc{Steve} and \textsc{People Tool}.}
\label{fig:appendix-short-queries}
\end{figure}
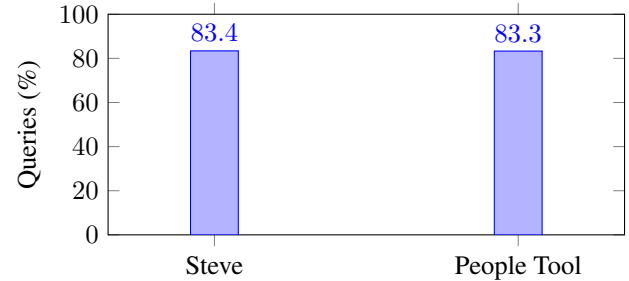

Query forms for the two systems are depicted in Figures~\ref{fig:appendix-query-length} and~\ref{fig:appendix-short-queries}. These findings provide evidence for the claims made in the paper regarding how the users carried forward their legacy behavior in terms of keyword-based searching despite having access to a GenAI-powered search engine. It is noteworthy that there was not much of an increase in query length in \textsc{People Tool} and short queries consisting of just one or two tokens were the norm in both cases.

Figure~\ref{fig:appendix-short-queries} illustrates the same trend from a different perspective. The proportion of one- or two-token searches remained virtually unchanged for both systems. It is noteworthy that while GenAI-based searches might demand more precise phrasing, a significant number of people still searched with short, keyword-type searches.

\subsection{Result Availability and GenAI Exposure}

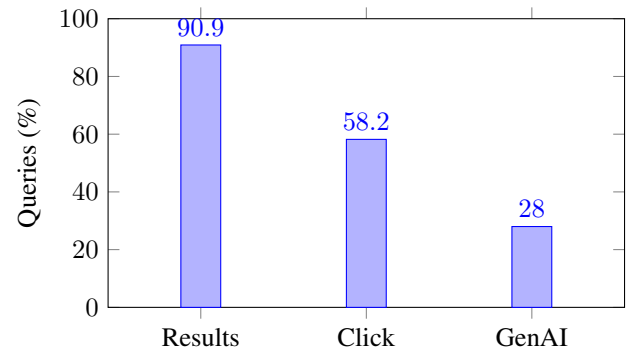
\begin{figure}[h]
\centering
\begin{tikzpicture}
\begin{axis}[
    ybar,
    width=\columnwidth,
    height=5.4cm,
    ymin=0,
    ymax=100,
    ylabel={Queries (\%)},
    symbolic x coords={Results,Click,GenAI},
    xtick=data,
    nodes near coords,
    nodes near coords align={vertical},
    bar width=15pt,
    enlarge x limits=.28
]
\addplot coordinates {
    (Results,90.9)
    (Click,58.2)
    (GenAI,28.0)
};
\end{axis}
\end{tikzpicture}
\caption{\textsc{People Tool} result availability, click behavior, and GenAI answer exposure.}
\label{fig:appendix-people-tool-funnel}
\end{figure}

Figure~\ref{fig:appendix-people-tool-funnel} distinguishes between three types of interaction phases in \textsc{People Tool}, which include whether there was a response from the search query, whether one of the responses was clicked on, and whether there was a response generated by GenAI. The reason for this distinction is that the use of \textsc{People Tool} does not automatically imply the use of GenAI.

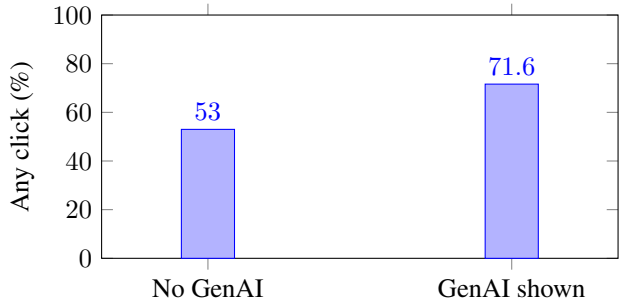
\begin{figure}[h]
\centering
\begin{tikzpicture}
\begin{axis}[
    ybar,
    width=\columnwidth,
    height=4.8cm,
    ymin=0,
    ymax=100,
    ylabel={Any click (\%)},
    symbolic x coords={No GenAI,GenAI shown},
    xtick=data,
    nodes near coords,
    nodes near coords align={vertical},
    bar width=20pt,
    enlarge x limits=.35
]
\addplot coordinates {
    (No GenAI,53.0)
    (GenAI shown,71.6)
};
\end{axis}
\end{tikzpicture}
\caption{Click behavior by GenAI answer exposure in \textsc{People Tool}.}
\label{fig:appendix-click-by-genai}
\end{figure}

Figure~\ref{fig:appendix-click-by-genai} compares click behavior when a GenAI answer was displayed and when it was not. Searches with a displayed GenAI answer had higher click-through than searches without one. This should be interpreted only as behavioral engagement. It does not show whether users trusted the answer, found it useful, or acted on it.

\subsection{Country and Test-Context Variation}

Table~\ref{tab:appendix-country-metrics} summarizes \textsc{People Tool} log indicators by country or test context. These differences help contextualize the theme of situated fit. However, they should not be interpreted as cultural differences by themselves. Variation may reflect local HR processes, rollout conditions, content availability, language, sample size, or task mix.

\begin{table}[h]
\centering
\caption{\textsc{People Tool} indicators by country or test context.}
\label{tab:appendix-country-metrics}
\scriptsize
\begin{tabularx}{\columnwidth}{p{0.7cm}rrrr}
\toprule
\textbf{Ctx.} & \textbf{N} & \textbf{Results} & \textbf{GenAI} & \textbf{Click} \\
 &  & \textbf{(\%)} & \textbf{(\%)} & \textbf{(\%)} \\
\midrule
EG & 205 & 85.4 & 6.8  & 49.3 \\
NL & 678 & 93.7 & 43.7 & 65.9 \\
PH & 26  & 92.3 & 42.3 & 61.5 \\
PK & 27  & 77.8 & 14.8 & 18.5 \\
QA & 59  & 94.9 & 10.2 & 57.6 \\
SA & 231 & 88.3 & 5.2  & 47.6 \\
\bottomrule
\end{tabularx}
\vspace{1mm}
{\raggedright\scriptsize \textit{Note:} Results = searches with returned results; GenAI = searches where a GenAI answer was displayed; Click = searches with any clicked result.\par}
\end{table}

Table~\ref{tab:appendix-country-intents} summarizes dominant search intents observed across country or test contexts. These patterns show that users were not bringing a single generic HR search need to the system. Instead, queries reflected local HR practices and task needs, including policy navigation, formal documents, benefits, insurance, and travel claims.

\begin{table}[h]
\centering
\caption{Dominant search intents by country or test context.}
\label{tab:appendix-country-intents}
\scriptsize
\begin{tabularx}{\columnwidth}{p{0.8cm}X}
\toprule
\textbf{Ctx.} & \textbf{Dominant observed search intents} \\
\midrule
NL & Policy and process navigation, including handbook, salary, leave, and Workday-related queries. \\
SA & Formal document and certificate-related queries, including letters and salary certificates. \\
QA & HR documents, benefits, medical, passport details, and employee profile queries. \\
EG & Benefits and insurance-related searches. \\
PK & Procedural travel-claim and reimbursement queries. \\
PH & Small sample of short, concrete queries with limited repeated patterns. \\
\bottomrule
\end{tabularx}
\vspace{1mm}
{\raggedright\scriptsize \textit{Note:} These patterns are descriptive. They should not be interpreted as direct evidence of national culture or user attitudes.\par}
\end{table}

\section{Descriptive Survey Analysis}
\label{app:survey-analysis}

This appendix reports descriptive survey information used to contextualize the interview findings. The survey was exploratory and was completed by 25 employees with access to \textsc{People Tool} and its GenAI search feature. The results should not be interpreted as statistically representative of the wider organization. Instead, they provide descriptive support for the qualitative analysis by showing patterns in reported use, perceived ease, helpfulness, trust, source-checking, and responses to poor GenAI answers.


\begin{table}[h]
\centering
\caption{Overview of survey sections and item types.}
\label{tab:survey-instrument-overview}
\scriptsize
\begin{tabularx}{\columnwidth}{p{1.9cm}X}
\toprule
\textbf{Section} & \textbf{Survey items} \\
\midrule
Context & Reported work or test country, employment level, and tenure. \\
Usage patterns & Frequency of using \textsc{Steve} and \textsc{People Tool}; common search topics in each system. \\
System comparison & Ease of use, helpfulness, trust, unhelpful or irrelevant results, and system preference. \\
GenAI-specific use & Frequency of receiving a GenAI-generated answer; helpfulness of GenAI summaries; source-clicking behavior; response when GenAI does not provide a good answer. \\
Future use and improvements & Open-text suggestions, additional comments, and invitation to participate in follow-up interviews. \\
\bottomrule
\end{tabularx}
\end{table}

\subsection{Survey Respondent Context}

The survey sample included respondents from seven workplaces. Table~\ref{tab:survey-country} reports the normalized country or test-location distribution. These distributions are included for transparency only; the small and uneven subgroup sizes mean that country-level differences should not be interpreted as statistically reliable effects. Table~\ref{tab:survey-position} reports respondents' employment levels. The sample was skewed towards professional and specialist roles (See Limitations section). Table~\ref{tab:survey-tenure} reports tenure distribution. The sample included both newer and long-tenured employees, supporting descriptive interpretation of organizational familiarity, but not inferential subgroup claims.

\begin{table}[h]
\centering
\caption{Survey respondents by normalized work or test location.}
\label{tab:survey-country}
\scriptsize
\begin{tabularx}{\columnwidth}{Xrr}
\toprule
\textbf{Location} & \textbf{N} & \textbf{\%} \\
\midrule
Germany & 12 & 48 \\
United States & 6 & 24 \\
India & 3 & 12 \\
Austria & 1 & 4 \\
Switzerland & 1 & 4 \\
France & 1 & 4 \\
Mexico & 1 & 4 \\
\bottomrule
\end{tabularx}
\end{table}

\begin{table}[h]
\centering
\caption{Survey respondents by employment level.}
\label{tab:survey-position}
\scriptsize
\begin{tabularx}{\columnwidth}{Xrr}
\toprule
\textbf{Employment level} & \textbf{N} & \textbf{\%} \\
\midrule
Professional / Specialist & 16 & 64 \\
Intern / Entry-level employee & 2 & 8 \\
Contractor / Temporary staff & 2 & 8 \\
Team Lead / Manager & 2 & 8 \\
Senior Manager / Director / Executive & 1 & 4 \\
Other employment category. & 1 & 4 \\
Prefer not to say & 1 & 4 \\
\bottomrule
\end{tabularx}
\end{table}

\begin{table}[!h]
\centering
\caption{Survey respondents by tenure.}
\label{tab:survey-tenure}
\scriptsize
\begin{tabularx}{\columnwidth}{Xrr}
\toprule
\textbf{Tenure} & \textbf{N} & \textbf{\%} \\
\midrule
Less than 1 year & 2 & 8 \\
1--3 years & 9 & 36 \\
4--7 years & 6 & 24 \\
8--15 years & 4 & 16 \\
More than 15 years & 4 & 16 \\
\bottomrule
\end{tabularx}
\end{table}

\subsection{System Use and Preference}

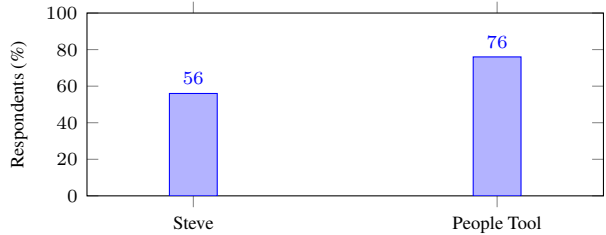
\begin{figure}[!h]
\centering
\begin{tikzpicture}
\begin{axis}[
    ybar,
    width=\columnwidth,
    height=4cm,
    ymin=0,
    ymax=100,
    ylabel={Respondents (\%)},
    symbolic x coords={Steve,People Tool},
    xtick=data,
    nodes near coords,
    nodes near coords style={font=\scriptsize},
    xticklabel style={font=\scriptsize},
    yticklabel style={font=\scriptsize},
    ylabel style={font=\scriptsize},
    bar width=18pt,
    enlarge x limits=.35
]
\addplot coordinates {(Steve,56) (People Tool,76)};
\end{axis}
\end{tikzpicture}
\caption{Respondents' weekly or daily use of each system.}
\label{fig:survey-system-use}
\end{figure}

\begin{figure}[!h]
\centering
\begin{tikzpicture}
\begin{axis}[
    xbar,
    width=6cm,
    height=4cm,
    xmin=0,
    xmax=70,
    xlabel={Respondents (\%)},
    symbolic y coords={Mixed / unclear,Prefer\textsc{ Steve},Prefer\textsc{ People Tool}},
    ytick=data,
    nodes near coords align={horizontal},        
    nodes near coords style={font=\scriptsize}, 
    xticklabel style={font=\scriptsize},
    yticklabel style={font=\scriptsize},
    xlabel style={font=\scriptsize},
    bar width=15pt,
    enlarge y limits=.30
]
\addplot coordinates {
    (28,Mixed / unclear)
    (12,Prefer\textsc{ Steve})
    (56,Prefer\textsc{ People Tool})
};
\end{axis}
\end{tikzpicture}
\caption{System preference reported in the survey.}
\label{fig:survey-preference}
\end{figure}
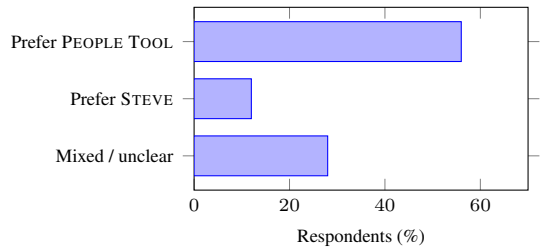

Figure~\ref{fig:survey-system-use} summarizes the reported frequency of use and system preference. \textsc{People Tool} was used weekly or daily by more respondents than \textsc{Steve}, but the continued use of both systems supports the interpretation that adoption was selective rather than a clean replacement. Figure~\ref{fig:survey-preference} shows that a majority clearly preferred \textsc{People Tool}, but a substantial minority preferred \textsc{Steve} or gave mixed or unclear preferences. This supports the paper's framing of parallel-system use and selective adoption.

\subsection{Perceived Ease, Helpfulness, and Trust}

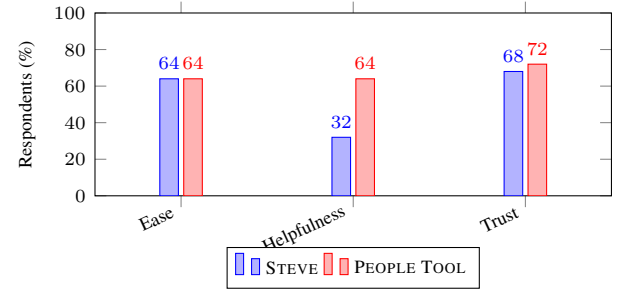
\begin{figure}[h]
\centering
\begin{tikzpicture}
\begin{axis}[
    ybar,
    width=\columnwidth,
    height=4cm,
    ymin=0,
    ymax=100,
    ylabel={Respondents (\%)},
    symbolic x coords={Ease,Helpfulness,Trust},
    xtick=data,
    x tick label style={font=\scriptsize, rotate=25, anchor=east},
    yticklabel style={font=\scriptsize},
    ylabel style={font=\scriptsize},
    legend style={font=\scriptsize, at={(0.5,-0.28)}, anchor=north, legend columns=2},
    nodes near coords,
    nodes near coords style={font=\scriptsize},
    bar width=7pt,
    enlarge x limits=.25
]
\addplot coordinates {(Ease,64) (Helpfulness,32) (Trust,68)};
\addplot coordinates {(Ease,64) (Helpfulness,64) (Trust,72)};
\legend{\textsc{Steve}, \textsc{People Tool}}
\end{axis}
\end{tikzpicture}
\caption{Survey ratings for ease, helpfulness, and trust. Ease = easy or very easy; helpfulness = very or extremely helpful; trust = mostly or completely trusted.}
\label{fig:survey-ratings}
\end{figure}

Figure~\ref{fig:survey-ratings} compares three system ratings. The key pattern is that \textsc{People Tool} was perceived as more helpful, while ease of use and trust changed less substantially. This supports the interpretation that perceived usefulness alone did not remove the need for trust calibration, search literacy, and organizational support.

\subsection{GenAI Answer Exposure, Helpfulness, and Verification}

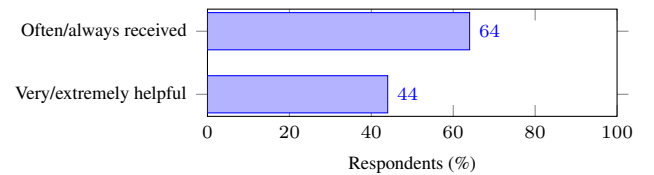
\begin{figure}[!h]
\centering
\begin{tikzpicture}
\begin{axis}[
    xbar,
    width= 7 cm,
    height=3cm,
    xmin=0,
    xmax=100,
    xlabel={Respondents (\%)},
    symbolic y coords={Very/extremely helpful,Often/always received},
    ytick=data,
    nodes near coords,
    nodes near coords style={font=\scriptsize},
    xticklabel style={font=\scriptsize},
    yticklabel style={font=\scriptsize},
    xlabel style={font=\scriptsize},
    bar width=14pt,
    enlarge y limits=.35
]
\addplot coordinates {
    (44,Very/extremely helpful)
    (64,Often/always received)
};
\end{axis}
\end{tikzpicture}
\caption{Reported GenAI answer exposure and summary helpfulness in \textsc{People Tool}.}
\label{fig:survey-genai-summary}
\end{figure}

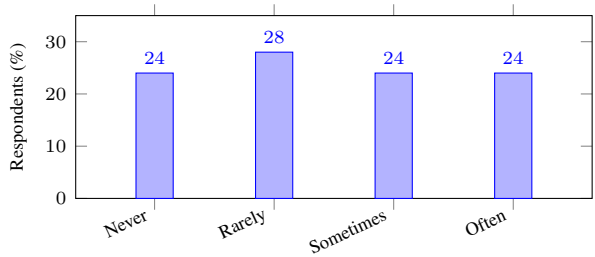
\begin{figure}[h]
\centering
\begin{tikzpicture}
\begin{axis}[
    ybar,
    width=\columnwidth,
    height=4cm,
    ymin=0,
    ymax=35,
    ylabel={Respondents (\%)},
    symbolic x coords={Never,Rarely,Sometimes,Often},
    xtick=data,
    nodes near coords,
    nodes near coords style={font=\scriptsize},
    xticklabel style={font=\scriptsize, rotate=25, anchor=east},
    yticklabel style={font=\scriptsize},
    ylabel style={font=\scriptsize},
    bar width=14pt,
    enlarge x limits=.22
]
\addplot coordinates {
    (Never,24)
    (Rarely,28)
    (Sometimes,24)
    (Often,24)
};
\end{axis}
\end{tikzpicture}
\caption{Frequency of clicking sources displayed in GenAI-generated answers.}
\label{fig:survey-source-clicking}
\end{figure}

Figure~\ref{fig:survey-genai-summary} summarizes respondents' reported experience of GenAI answers in \textsc{People Tool}. Most respondents reported receiving GenAI answers often or always, but fewer rated GenAI summaries as very or extremely helpful. Figure~\ref{fig:survey-source-clicking} shows how often respondents clicked sources displayed in GenAI-generated answers. Source-checking was divided: 52\% never or rarely clicked sources, while 48\% clicked them sometimes or often. This should be interpreted as uneven verification behavior, not as a direct measure of trust.

\subsection{Responses to Poor GenAI Answers}

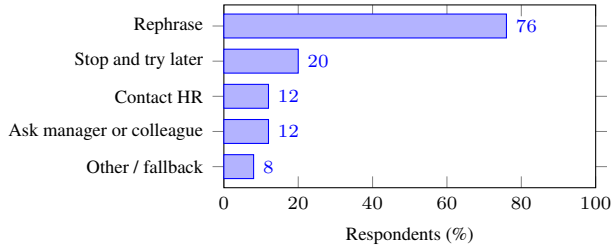
\begin{figure}[!h]
\centering
\begin{tikzpicture}
\begin{axis}[
    xbar,
    width= 6.5 cm,
    height=4cm,
    xmin=0,
    xmax=100,
    xlabel={Respondents (\%)},
    symbolic y coords={Other / fallback,Ask manager or colleague,Contact HR,Stop and try later,Rephrase},
    ytick=data,
    nodes near coords,
    nodes near coords style={font=\scriptsize},
    xticklabel style={font=\scriptsize},
    yticklabel style={font=\scriptsize},
    xlabel style={font=\scriptsize},
    bar width=9pt,
    enlarge y limits=.15
]
\addplot coordinates {
    (76,Rephrase)
    (20,Stop and try later)
    (12,Contact HR)
    (12,Ask manager or colleague)
    (8,Other / fallback)
};
\end{axis}
\end{tikzpicture}
\caption{Reported responses when GenAI did not provide a good answer. Responses were multi-select, so percentages do not sum to 100.}
\label{fig:survey-genai-repair}
\end{figure}

Figure~\ref{fig:survey-genai-repair} reports what respondents said they would do when GenAI did not provide a good answer. Rephrasing was the most common response, indicating that users actively repair failed interactions by changing how they ask. Escalation to human or other information channels was less frequent but remains important for understanding fallback practices.

\subsection{Common Search Topics}

Table~\ref{tab:survey-search-topics} reports the most common standardized search topics selected by respondents for \textsc{Steve} and \textsc{People Tool}. These topic distributions show that respondents used both systems for practical HR needs, including policies, leave, payroll, sickness, travel, overtime, and benefits. Because respondents could select multiple topics, counts are not mutually exclusive.

\begin{table}[h]
\centering
\caption{Top standardized search topics reported for each system.}
\label{tab:survey-search-topics}
\scriptsize
\begin{tabularx}{\columnwidth}{Xrr}
\toprule
\textbf{Search topic} & \textbf{Steve} & \textbf{People Tool} \\
\midrule
Employee handbook / HR policies & 13 & 14 \\
Annual leave / holiday policies & 8 & 10 \\
Salary slip / payroll information & 10 & 7 \\
Sick leave requests / policies & 4 & 7 \\
Travel expenses / reimbursements & 6 & 5 \\
Overtime policy / work hours & 6 & 5 \\
Bonus / incentives information & 5 & 5 \\
Maternity or paternity leave & 5 & 4 \\
IT access & 3 & 4 \\
\bottomrule
\end{tabularx}
\vspace{1mm}
{\raggedright\scriptsize \textit{Note:} Topic responses were multi-select. Counts therefore indicate how many respondents selected each topic and may sum to more than the number of respondents.\par}
\end{table}


Table~\ref{tab:survey-instrument-overview} summarizes the structure of the survey instrument. The questionnaire compared the legacy system, \textsc{Steve}, with the newer GenAI-enhanced platform, \textsc{People Tool}, and included both closed-ended and open-text questions. GenAI-specific questions asked respondents about answer exposure, summary helpfulness, source-clicking, and response strategies when the GenAI answer was not useful.

\begin{table*}[p]
\centering
\caption{Anonymized survey instrument.}
\label{tab:survey-instrument}

\begin{tcolorbox}[
    enhanced,
    colback=white,
    colframe=black!35,
    boxrule=0.6pt,
    arc=1.5mm,
    left=2mm,
    right=2mm,
    top=1.5mm,
    bottom=1.5mm,
    title=\textbf{Survey instrument: Comparing \legacy{} and \newtool{} GenAI Search},
    fonttitle=\bfseries,
    coltitle=black,
    colbacktitle=black!6,
    attach boxed title to top left={xshift=2mm,yshift=-1.5mm},
    boxed title style={
        colframe=black!20,
        boxrule=0.4pt,
        arc=1mm,
        left=1.5mm,
        right=1.5mm,
        top=0.8mm,
        bottom=0.8mm
    }
]

\scriptsize
\renewcommand{\arraystretch}{1.16}
\setlength{\tabcolsep}{5pt}

\begin{tabularx}{\textwidth}{@{}p{0.18\textwidth}p{0.09\textwidth}Y@{}}
\toprule
\textbf{Survey section} & \textbf{Items} & \textbf{Questions and response format} \\
\midrule

\textbf{Survey context and system overview}
& Intro
& The survey introduced the study as research on a GenAI-supported HR search feature in a large multinational organization. It explained that responses were anonymous, used for research and improvement, and expected to take approximately 10 minutes. 
The original researcher's name, institution, company name, department, email address, form URL, and screenshots were removed in this anonymized version.

\begin{surveyitems}
    \item \textbf{\legacy:} legacy HR query-based search and database system, matching employee questions with pre-scripted answers and HR documents.
    \item \textbf{\newtool:} GenAI-enhanced HR knowledge and support hub, offering source-backed answers based on human-reviewed organizational content.
\end{surveyitems}
\\

\addlinespace[2pt]
\textbf{I. Context and demographics}
& 1--3
& \begin{surveyitems}
    \item \textbf{1.} Work or test country. \emph{Open text.}
    \item \textbf{2.} Current position or employment level. \emph{Options:} intern / entry-level employee; professional / specialist; team lead / manager; senior manager / director / executive; contractor / temporary staff; prefer not to say; other.
    \item \textbf{3.} Organizational tenure. \emph{Options:} less than 1 year; 1--3 years; 4--7 years; 8--15 years; more than 15 years; prefer not to say.
\end{surveyitems}
\\

\addlinespace[2pt]
\textbf{II. Usage patterns and system comparison}
& 4--15
& \begin{surveyitems}
    \item \textbf{4.} Frequency of using \legacy{} and \newtool. \emph{Scale:} never; rarely (less than monthly); occasionally (monthly); frequently (weekly); very frequently (daily).
    \item \textbf{5--6.} HR topics searched most often in \legacy{} and \newtool. \emph{Multi-select options:} employee handbook / HR policies; annual leave and holiday policies; salary slip / payroll information; sick leave requests and policies; travel expense and reimbursements; maternity or paternity leave; overtime policy and work hours; bonus / incentives information; IT access; other.
    \item \textbf{7.} Ease of use for \legacy{} and \newtool. \emph{Scale:} very difficult; difficult; neutral; easy; very easy.
    \item \textbf{8.} Difficulties experienced using either system. \emph{Open text.}
    \item \textbf{9.} Helpfulness of \legacy{} and \newtool{} for completing tasks. \emph{Scale:} not helpful at all; slightly helpful; moderately helpful; very helpful; extremely helpful.
    \item \textbf{10.} Elaboration on unhelpful experiences. \emph{Open text.}
    \item \textbf{11.} Trust in search results from \legacy{} and \newtool. \emph{Scale:} do not trust at all; slightly trust; moderately trust; mostly trust; trust completely.
    \item \textbf{12.} Elaboration on lack of trust. \emph{Open text.}
    \item \textbf{13.} Frequency of unhelpful or irrelevant answers. \emph{Scale:} never (less than 1/10); rarely (1/10); sometimes (3/10); half the time (5/10); often (7/10); almost always (9/10).
    \item \textbf{14.} Topics that returned unhelpful or irrelevant results. \emph{Open text.}
    \item \textbf{15.} Preferred system and reason. \emph{Open text.}
\end{surveyitems}
\\

\addlinespace[2pt]
\textbf{III. \newtool{} GenAI search-specific questions}
& 16--19
& \begin{surveyitems}
    \item \textbf{16.} Frequency of receiving a GenAI-generated answer when searching in \newtool. \emph{Scale:} always; often; sometimes; rarely; never.
    \item \textbf{17.} Helpfulness of GenAI summaries in \newtool. \emph{Scale:} extremely helpful; very helpful; somewhat helpful; slightly helpful; not helpful at all.
    \item \textbf{18.} Frequency of clicking sources displayed in GenAI-generated answers. \emph{Scale:} always; often; sometimes; rarely; never.
    \item \textbf{19.} Response when a GenAI-generated answer is not good. \emph{Multi-select options:} I always get good GenAI-generated answers; rephrase the question; stop and try again later; contact HR directly; ask my manager or colleague; other.
\end{surveyitems}
\\

\addlinespace[2pt]
\textbf{IV. Future use and improvements}
& 20--22
& \begin{surveyitems}
    \item \textbf{20.} Top two or three things that would make \newtool{} more useful. \emph{Open text.}
    \item \textbf{21.} Additional comments. \emph{Open text.}
    \item \textbf{22.} Willingness to join a short confidential follow-up interview; contact information stored separately and excluded from analysis. \emph{Open text.}
\end{surveyitems}
\\

\bottomrule
\end{tabularx}

\vspace{1mm}
\footnotesize
\emph{Note.} This table reports the anonymized survey instrument. Original organization names, system names, researcher names, institutional affiliations, email addresses, form URLs, and screenshots were removed or replaced with anonymized labels.

\end{tcolorbox}
\end{table*}

\section{Interview Study Materials}
\label{app:interview-materials}

This appendix provides the interview materials used in the qualitative phase of the study. Semi-structured interviews were the primary interpretive data source and were used to examine how employees made sense of \textsc{Steve}, \textsc{People Tool}, and the GenAI search feature in practice. The interview guide was designed to move from participants' general background and transition experiences to more focused questions about usefulness, trust, adoption dynamics, and organizational support.

The interview guide shown below is the English version used in the main study. Minor adjustments in wording and follow-up prompts were made during interviews to maintain conversational flow and to probe relevant examples in greater depth, but the overall thematic structure remained stable across participants.

\begin{table*}[t]
\centering
\caption{Semi-structured interview guide.}
\label{tab:interview-guide}

\begin{tcolorbox}[
    enhanced,
    colback=white,
    colframe=black!35,
    boxrule=0.6pt,
    arc=1.5mm,
    left=2mm,
    right=2mm,
    top=1.5mm,
    bottom=1.5mm,
    title=\textbf{Interview guideline v.02 (English)},
    fonttitle=\bfseries,
    coltitle=black,
    colbacktitle=black!6,
    attach boxed title to top left={xshift=2mm,yshift=-1.5mm},
    boxed title style={
        colframe=black!20,
        boxrule=0.4pt,
        arc=1mm,
        left=1.5mm,
        right=1.5mm,
        top=0.8mm,
        bottom=0.8mm
    }
]

\small
\renewcommand{\arraystretch}{1.18}
\setlength{\tabcolsep}{5pt}

\begin{tabularx}{\textwidth}{@{}p{0.22\textwidth}p{0.11\textwidth}Y@{}}
\toprule
\textbf{Interview section} & \textbf{Approx. time} & \textbf{Guiding prompts} \\
\midrule

\textbf{I. Warm-up} 
& 2--3 min
& \begin{guideitems}
    \item Role and tenure.
    \item Comfort level with digital tools at work.
\end{guideitems}
\\

\addlinespace[2pt]
\textbf{II. Transition experiences: \textsc{Steve} $\rightarrow$ \textsc{People Tool}}
& 5--6 min
& \begin{guideitems}
    \item Prior experience with \textsc{Steve}.
    \item Experience with \textsc{People Tool} and perceived differences.
    \item Moments where one system felt easier or more useful than the other.
\end{guideitems}
\\

\addlinespace[2pt]
\textbf{III. Usefulness and limitations of GenAI}
& 6 min
& \begin{guideitems}
    \item Situations where \textsc{People Tool} was particularly helpful.
    \item Situations where \textsc{People Tool} was unhelpful and how the participant responded.
    \item Types of HR questions best suited to \textsc{People Tool} versus human support, such as colleagues or HR.
\end{guideitems}
\\

\addlinespace[2pt]
\textbf{IV. Trust and information behaviors}
& 4--5 min
& \begin{guideitems}
    \item Factors influencing trust or hesitation in \textsc{People Tool} responses.
    \item Whether participants checked source articles or relied on summaries, and why.
\end{guideitems}
\\

\addlinespace[2pt]
\textbf{V. Adoption dynamics and social context}
& 6--7 min
& \begin{guideitems}
    \item Observed differences in adoption by tenure, role, or country.
    \item Influence of managers, HR, or colleagues on \textsc{People Tool} use.
    \item Smoothness of the \textsc{Steve} $\rightarrow$ \textsc{People Tool} transition.
    \item Fit of \textsc{People Tool} for shop-floor versus office employees.
    \item Desired organizational support or training to ease the transition.
\end{guideitems}
\\

\addlinespace[2pt]
\textbf{VI. Closing reflections}
& 4--5 min
& \begin{guideitems}
    \item Key factor influencing adoption or avoidance of \textsc{People Tool}.
    \item Message to organizational leadership about employees' experiences with AI in HR.
\end{guideitems}
\\

\bottomrule
\end{tabularx}

\vspace{1mm}
\footnotesize
\emph{Note.} The guide was used flexibly during semi-structured interviews; follow-up questions were adapted to participants' roles, examples, and prior responses.

\end{tcolorbox}
\end{table*}

\end{document}